# Mid-Band Dissipative Spatial Solitons

**Kestutis Staliunas**

*PTB Braunschweig, Bundesallee 100, 38116 Braunschweig, Germany*

*kestutis.staliunas@ptb.de*

**Abstract**

We show dissipative spatial solitons in nonlinear optical micro-resonators in which the refractive index is laterally modulated. In addition to "normal" and "staggered" dissipative solitons, similar to those in spatially modulated conservative systems, a narrow "mid-band" soliton is shown, having no counterparts in conservative systems.

Dissipative spatial solitons (DSS) have been theoretically predicted in bistable nonlinear optical resonators in general [1], and in particular: in lasers with saturable absorber [2], in two-photon lasers [3] in optical parametric oscillators [4], in semiconductor micro-resonators [5]. In some systems the DSS have been demonstrated experimentally: in lasers with saturable absorbers [6], in four wave mixing [7], and in semiconductor micro-resonators [8]. The DSS, being bistable, can be potentially applied for parallel information processing and storage [9].

One of the fundamental limitations of DSS is their spatial size, evaluated as $x_0 \propto \sqrt{L\lambda Q}$, where $\lambda$ is the wavelength of the radiation, $L$ and $Q$ are the full length and finesse of the resonator. Even in the smallest soliton supporting systems, the semiconductor resonators, with typical parameter values: $\lambda \approx 0.5 \mu m$ (in material with refraction index $\approx 2$), $L \approx 5 mm$, and $Q \approx 100$, the width of the soliton is $x_0 \approx 15 \mu m$, as follows from the above evaluation, and from recent experimental observations [8].

The above evaluation is based on the corresponding dynamical equations, where the spatial coordinate of the solution is scaled by the square root of the diffraction coefficient $d = L\lambda Q/(4\pi)$. This evaluation also follows from the linear theory of the resonator, stating that a spatial light structure of minimal complexity (e.g. a vortex or DSS) requires that the Fresnel number of resonator $F = \pi x_0^2/(L\lambda Q)$ is of order of one or more. The above limitation of the minimum size of the DSS is a fundamental one, occurring due to the diffraction of light in the resonator.

We show in this letter that the above limitation can be overcome for DSS in a resonator with a lateral periodic modulation of parameters, such as refractive index of material, the resonator mirror surface, the gain coefficient, or others.

Conservative spatial solitons in periodically modulated nonlinear materials (nonlinear photonic crystals) have recently gained popularity [10]. Discrete Kerr-type solitons in materials with $\chi^{(3)}$ nonlinearity [11], as well as parametric discrete solitons in materials with $\chi^{(2)}$ nonlinearity [12] have been recently shown. Those are the "normal" solitons with the carrier spatial wave-number being around zero (the analogs of the spatial solitons in homogeneous nonlinear media), also the "band-edge", or "staggered" solitons, with the carrier spatial wave-number being around half of the modulation wave-number. The field phase changes its value by $\pi$ between the modulation fringes across the staggered soliton. The staggered solitons have no counterpart in homogeneous media, and the effective diffraction for the staggered solitons has a negative value.

We show here the existence of normal and staggered DSS in nonlinear resonators with laterally modulated parameters. These DSS have the above described analogs in nonlinear photonic crystals (in conservative systems). We also find that dissipative nonlinear resonators support another type of the solitons: the DSS with the carrier spatial frequency between those for normal and for staggered solitons. Indeed, if the effective diffraction changes its sign from positive for normal solitons, to negative for staggered solitons, then there should be a point where the effective diffraction is zero. The spatial spectra of these soliton are centered at the zero-diffraction point, i.e. located close to the middle of the band of extended waves. Therefore we name them "mid-band" solitons. Since the effective diffraction vanishes for the mid-band DSS, their size is not limited by diffraction, i.e. by: $x_0 \propto \sqrt{L\lambda Q}$. The width of the mid-band DSS then depends on the next higher diffraction order. Consequently the mid-band DSS can be much narrower than above evaluated, and can be reduced down to one wavelength.

We show first the DSS (normal, staggered and mid-band ones) by numerical integration of the equations for a laser with a saturable absorber, which is perhaps mathematically the simplest system supporting DSS. Next we evaluate the parameters of the solitons from the linear theory and compare with the numerical results. Also we prove that the effective diffraction for a mid-band DSS is eliminated, by showing that the soliton width is independent of the strength of nonlinear Kerr-focusing term introduced additionally in the equation.

As a model system a laser with saturable absorber was numerically investigated. It is known, that this system supports DSSs [2]. We note, however, that the principal results are also valid for solitons in other nonlinear optical resonators. The laser with saturable absorber in paraxial- and mean field approximations is described by:

$$\frac{\partial A(\vec{r},t)}{\partial t} = \left( \frac{D_0}{1+|A|^2} - 1 - \frac{\alpha_0}{1+|A|^2/I_a} + id\nabla^2 + g\nabla^2 + iV(\vec{r}) \right) A(\vec{r},t). \qquad (1)$$

The right hand side contains the terms of saturating gain, linear losses, saturating losses, diffraction, diffusion, and spatially varying refractive index: $V(\vec{r}) = m(e^{iqx} + e^{-iqx})$, respectively. $D_0$ is the unsaturated gain, $\alpha_0$ is the unsaturated absorption, $I_a$ is the saturation intensity of the absorber, $d = L\lambda Q/(4\pi)$ is the diffraction coefficient, $g$ is the diffusion coefficient (usually $g \ll d$), and $m$ and $q$ are the amplitude and the wave-number of the refractive index modulation in the transverse direction of resonator. See e.g. also [3] for detailed description of the model (1).

The diffusion term lumps together the material diffusion (e.g. diffusion of population inversion), the limited width of the gain line, as well as the diffusion of light (i.e. spatial frequency filtering). In most systems diffusion can be considered small compared with diffraction (e.g. if the resonator mode resonance width is small compared with the gain line width). In the numerical integrations throughout the letter the diffusion/diffraction ratio is fixed to $g/d = 10^{-4}$, but the results depend only negligibly on this ratio.

We investigate a one dimensional system i.e. one transverse dimension plus time. The system was numerically integrated using a split-step technique by changing from the space domain to the spatial Fourier domain in every integration step, which imposes periodic boundary conditions. Fig.1 shows the typical spatial profiles (left) and spatial power spectra (right) of stable dissipative solitons.

a) Normal soliton. The soliton envelope is modulated due to the spatial modulation of the refractive index. The power spectrum consists of a central component (spatial carrier frequency), and progressively decreasing sidebands. We filtered out the sidebands from the spatial spectrum, and, using inverse Fourier transform, recovered the unmodulated component of the soliton (shown by a dashed line). The normal DSS is stable for the range of pump parameter $1.59 < D_0 < 1.70$.

b) Staggered soliton. The field phase changes by $p$ between the neighboring modulation fringes, therefore the field amplitude is modulated by 100%. The power spectrum contains two strong components with the spatial frequencies at $k_0 \approx \pm q/2$. Evidently the staggered soliton is wider than the normal one. The staggered soliton is stable for the range of pump parameter $1.66 < D_0 < 1.76$.

3) Mid-band soliton. The power spectrum contains a strong component centered at the spatial frequency of $0 < k_0 < q/2$. The soliton is much narrower than the normal and the staggered solitons, and is stable for a range of pump parameter $1.68 < D_0 < 1,79$. Differently from two previous cases the mid-band soliton is never stationary, but moves with a constant velocity (evaluated below).

The soliton stability range and the form of the soliton envelope depends on the nonlinear properties of the system. However, in order to evaluate roughly the width of the DSS, the linear analysis is useful. Thus we consider two spatial harmonics of the field: $A(x,t) = e^{ik_0 x}\left(A_0(t) + A_1(t)e^{-iqx}\right)$, where the field components with the wave-vectors $k_0$ and $k_1 = k_0 - q$ (closest to zero) are considered. The linear propagation of waves in the resonator

(equation (1) keeping only the linear in field amplitude and conservative terms) results in a system of coupled equations for the field components:

$$\partial A_0/\partial t = -idk_0^2 A_0 + imA_1 \qquad (2.a)$$

$$\partial A_1/\partial t = -idk_1^2 A_1 + imA_0 \qquad (2.b)$$

The solution $\partial A_{0,1}/\partial t = i\mathbf{w}A_{0,1}$ of (2) results in an oscillation frequency:

$$\mathbf{w} = \frac{dq^2}{2}\left(2k - 2k^2 - 1 \pm \sqrt{f^2 + (1-2k)^2}\right) \qquad (3)$$

where the parameter $f = 2m/(dq^2)$ has a meaning of a modulation depth as induced by the spatially periodic perturbation, i.e. of a contrast of modulation fringes. $k = k_0/q$ is the carrier spatial wave-number normalized to the spatial wave-number of the modulation. Fig.2.a. shows the solution (3), illustrating the appearance of a band-gap for nonzero modulation amplitude $m$. The effective diffraction coefficient for the system of coupled waves $d_{eff} = -1/2\,\partial^2 \mathbf{w}/\partial k_0^2$ is:

$$d_{eff} = d\left(1 - f^2/(f^2 + (1-2k)^2)^{3/2}\right) \qquad (4)$$

Fig.2.b. shows the solution (4), illustrating the appearance of a zero-diffraction point, at which the spectrum of mid-band soliton is centered. Next we calculate the effective diffraction coefficients, and evaluate the parameters of the soliton in the limit of small induced modulation $a \ll 1$ (we note that the modulation parameter $m$ must not be necessarily small for this limit).

1) For the waves along the optical axis of the resonator, corresponding to the normal DSS with the carrier spatial wave-number $k = 0$: $d_{eff} = d(1 - f^2 + ...)$. The width of the soliton scales as $x_0 \propto \sqrt{d_{eff}} = \sqrt{d}(1 - f^2/2 + ...)$.

2) For the waves corresponding to the band-edge, corresponding to the staggered DSS with the carrier spatial wave-number $k = 1/2$: $d_{eff} = d(1 - 1/f)$, is negative, as expected for band-edge solitons. The width of the staggered soliton scales as $x_0 \propto \sqrt{|d_{eff}|} = \sqrt{d/f}$. Our numerical calculations (not shown) correspond well with the above scaling.

3) The effective diffraction is zero at a wavenumber: $k = \left(1 - \sqrt{f^{4/3} - f^2}\right)/2$, as follows from (4). The size of the mid-band DSS can be evaluated from the next higher order diffraction $d_{eff,3} = 1/6 \partial^3 w/\partial k_0^3$ : $d_{eff,3} = 2\sqrt{1-f^{2/3}} d^{5/3} q^{1/3}/(2m)^{2/3}$. The soliton width then scales as $x_0 \propto |d_{eff,3}|^{1/3}$.

In first two cases the narrowing of the normal and staggered solitons depend solely on the contrast of modulation fringes $f$. The situation is different for the mid-band soliton, where the narrowing does not depend solely on the contrast $f$. The expression for $d_{eff,3}$ rewritten in a different way: $d_{eff,3} = 2\sqrt{1-f^{2/3}}/f^{2/3} \cdot d/q$ indicates, that the width of the soliton at a given contrast of modulation fringes $f$ decreases with increasing wave-number of modulation (and correspondingly increasing amplitude of modulation, to keep $f = 2m/(dq^2)$ constant). This means that one can achieve arbitrary narrowing of mid-band DSS even for small contrast of its modulation fringes.

The group velocity of the mid-band soliton, as following from (3), is: $v_{gr} = \partial w/\partial k_0 = qd(1-f^{2/3})^{3/2}$. Fig.3 shows the numerically calculated width and group velocity of the DSS depending on the modulation parameter. We depicted these values for the pump parameter at the boundaries of soliton stability range. Fig.3 indicates that the evaluation within the linear theory (solid lines) corresponds well with numerical results, at least for small contrast of modulation fringes (up to $f \leq 0.2$).

Finally we give a numerical proof that the effective diffraction vanishes for the mid-band DSS. For this purpose we modify the equation for the laser with saturable absorber by introducing phenomenologically an additional Kerr focussing/defocusing term:

$$\frac{\partial A(\vec{r},t)}{\partial t} = \left( \frac{D_0}{1+|A|^2} - 1 - \frac{a_0}{1+|A|^2/I_a} + id\nabla^2 + g\nabla^2 + iV(\vec{r}) + ic|A|^2 \right) A(\vec{r},t). \qquad (5)$$

In an experiment the presence of Kerr focusing/defocusing media in laser resonators, or the so called $a$-factor in semiconductor resonators results in a focusing/defocusing term in (5). Fig.4 shows the dependence of the width of the spatial solitons on parameter $c$. As expected, the normal DSS become narrower, and the staggered DSS become wider for increasing focusing. Remember that the effective diffraction is negative for staggered DSSs. However, the width (Fig.4), also the stability range (not shown) of mid-band DSS remain totally unaffected by focusing/defocusing, thus proving, that the effective diffraction for this kind of soliton is completely eliminated.

The numerical analysis including focusing/defocusing also indicates the robustness of mid-band DSSs with respect to structural modifications of the system. It allows to expect, that the mid-band solitons exist not only for lasers with saturable absorber, but represent a class of solitons which can be realized in a variety of nonlinear optical systems.

Concluding, we predict the mid-band soliton in dissipative, transversally modulated systems, and show that the soliton size is much smaller than that of conventional (normal and staggered) solitons. The size reduction is related with the elimination of the effective diffraction. We also show the robustness of the mid-band solitons with respect to structural modification of the system.

The letter focuses on DSS in one spatial dimension. The modification for two-dimensional case is straightforward, by considering square, or hexagonal modulation of parameters in two dimensional transverse plane.

The mid-band DSS move with a constant group velocity. This might be a difficulty for experimental observation of such solitons. One remedy against that difficulty is the use of moving refractive index gratings. Due to the Lorenz invariance of (1), and (5), the refractive index moving with a particular velocity would result in band-mid DSSs stationary in the transverse space.

The work has been supported by Sonderforschungsbereich 407 of Deutsche Forschungsgemeinschaft, and by Network PHASE of European Science Foundation. Discussions with C. O. Weiss, and R. Vilaseca are gratefully acknowledged.

**References**

1. D.W.McLaughlin, J.V.Moloney, and A.C.Newell, Phys. Rev. Letts, **51** 75 (1983); N.N.Rosanov, and G.V.Khodova, Opt. Spectrosc. **65**, 449 (1988); S.Fauve, and O.Thual, Phys. Rev. Lett. **64**, 282 (1990). M.Tlidi, and P.Mandel, and R.Lefever, Phys. Rev. Lett. **73**, 640 (1994).
2. N.N.Rosanov, JOSA B **7** 1057 (1990)
3. R. Vilaseca, M. C. Torrent, J. García-Ojalvo, M. Brambilla, and M. San Miguel, Phys. Rev. Lett. **87**, 083902 (2001)
4. K.Staliunas and V.J.Sanchez-Morcillo, Opt. Commun. **139,** 306 (1997); C.Etrich, U.Pechel, and F.Lederer, Phys. Rev. Lett. **79**, 2454 (1997); K.Staliunas, and V.J.Sanchez-Morcillo, Phys. Rev. **A57,** 1454 (1998);
5. D.Michaelis, U.Peschel, F.Lederer, Phys. Rev. **A56** R3366 (1997); M.Brambilla, L.A.Lugiato, F.Prati, L.Spinelli, and W.J.Firth, Phys. Rev. Lett. **79**, 2042 (1997)
6. V.B.Taranenko, K.Staliunas, and C.O.Weiss, Phys.Rev. **A56,** 1582 (1997); G.Slekys, K.Staliunas, C.O.Weiss, Opt. Commun. **149,** 113 (1998).
7. V.B.Taranenko, K.Staliunas, and C.O.Weiss, Phys.Rev.Letts, **81** 2236 (1998).
8. V.B.Taranenko, I.Ganne, R.J.Kuszelewicz, and C.O.Weiss Phys.Rev. **A61**, 063818 (2000); S.Barland, J.R.Tredicce, M.Brambilla, L.A.Lugiato, S.Balle, M.Giudici, T.Maggipinto, L.Spinelli, G.Tissoni, Th.Knödl, M.Miller, and R.Jäger, Nature **419**, 699 (2002);
9. G.S.McDonald, and W.J.Firth, J. Opt. Soc. Am. **B7**, 1328 (1990); W.J.Firth, A.J.Scroggie, Phys. Rev. Letts, **76** 1623 (1996).
10. D.N.Cristodoulides, and R.I.Joseph, Opt.Lett. **13**, 794 (1988); A.Aceves, C.De.Angeles, T.Pechel, R.Muschall, F.Lederer, S.Trillo, and S.Wabnitz, Phys.Rev. **E53,** 1172 (1996);
11. A.C.Scott, and L.Macneil, Phys.Lett A **98**, 87 (1983); S.Darmanyan, A.Kobyakov, F.Lederer, and L.Vazquez, Phys.Rev. **B59,** 5994 (1998);
12. A.Kobyakov, S.Darmanyan, T.Pertsch, and F.Lederer, Phys.Rev. **E57,** 2344 (1998); T.Pechel, U.Peschel, and F.Lederer, Phys.Rev. **E57,** 1127 (1998);

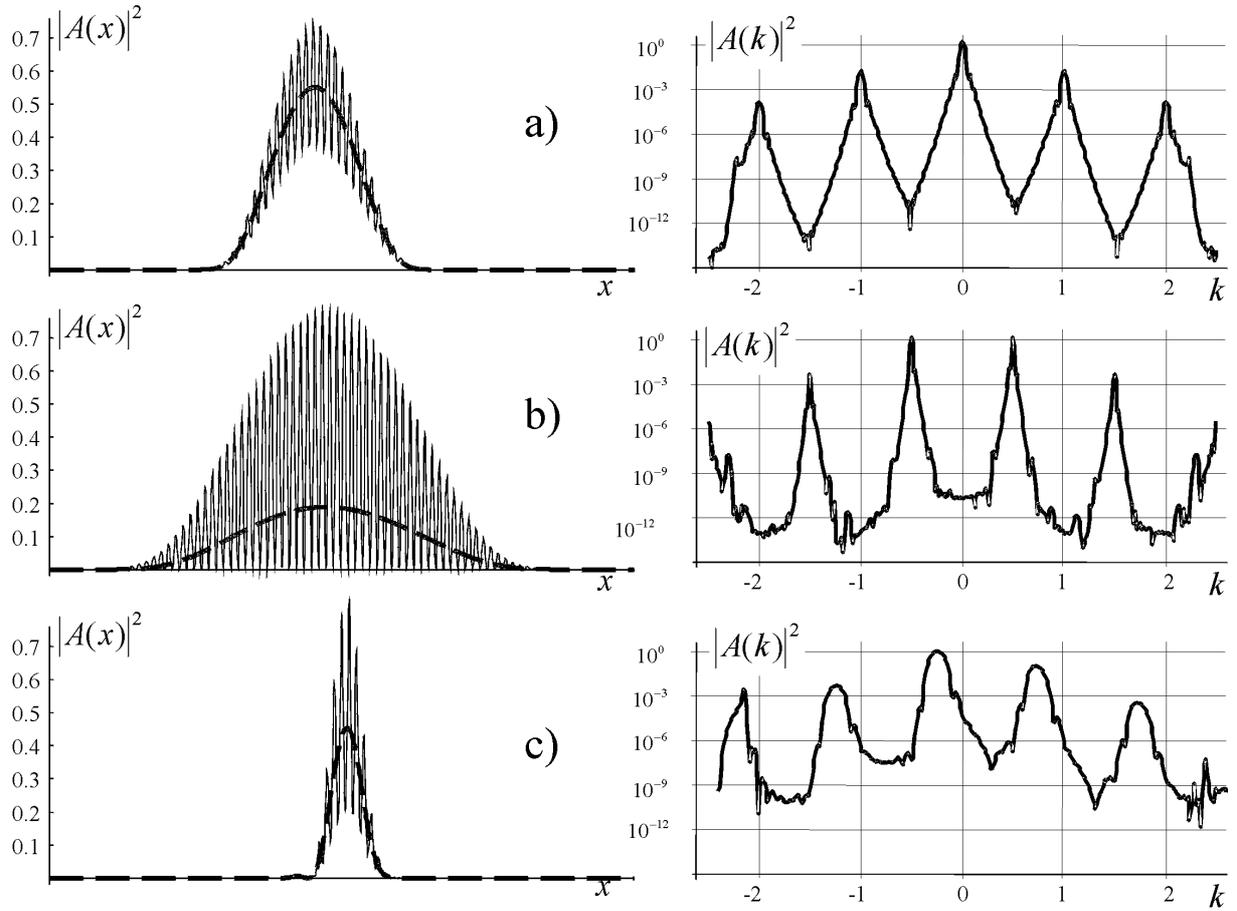

Fig.1. The envelopes (left) and the spatial power spectra (right) for normal (a), staggered (b), and mid-band (c) DSS, as obtained by numerical integration of (1). Dashed lines represent the envelopes of homogeneous component of the soliton (explanations in the text). The parameters are: $I_a = 0.01$, $a = 5$, $D_0 = 1.7$, $d = 0.001$ (the size of integration area is 1), $q = 80 * 2\pi$; $f = 2m/(dq^2) = 0.198$.

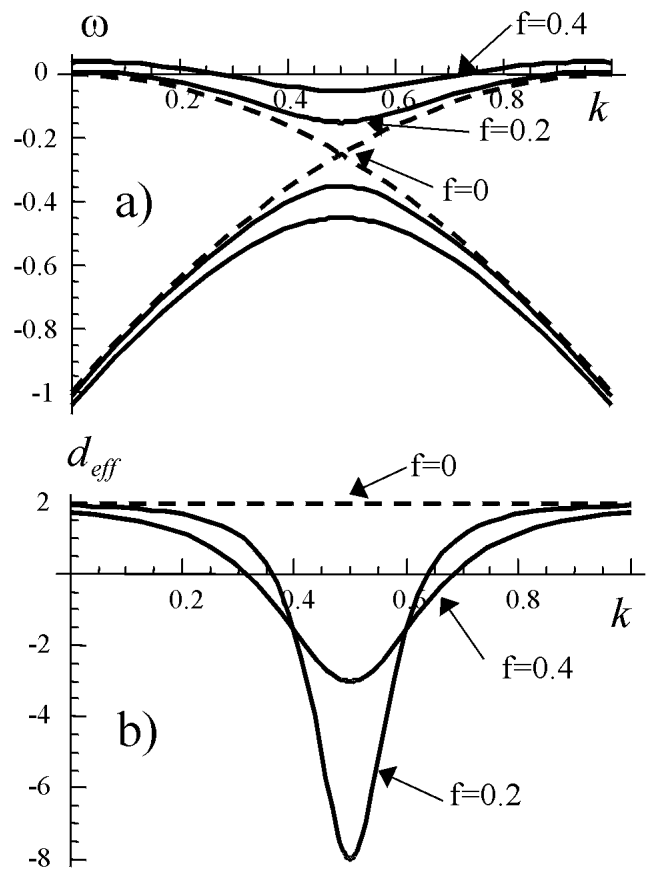

Fig.2. Frequency (a) as given by (3), and diffraction (b) as given by (4) of linearly coupled plane waves, depending on the normalized spatial carrier frequency $k = k_0/q$, for different coupling coefficients (modulation parameters): $m = 0$ dashed, and $m = 0.1$ ($f = 0.2$), $m = 0.2$ ($f = 0.4$) - solid lines. Other parameters are: $d = 1$, $q = 1$. The case $f = 0.2$ corresponds to parameters for Fig.1.

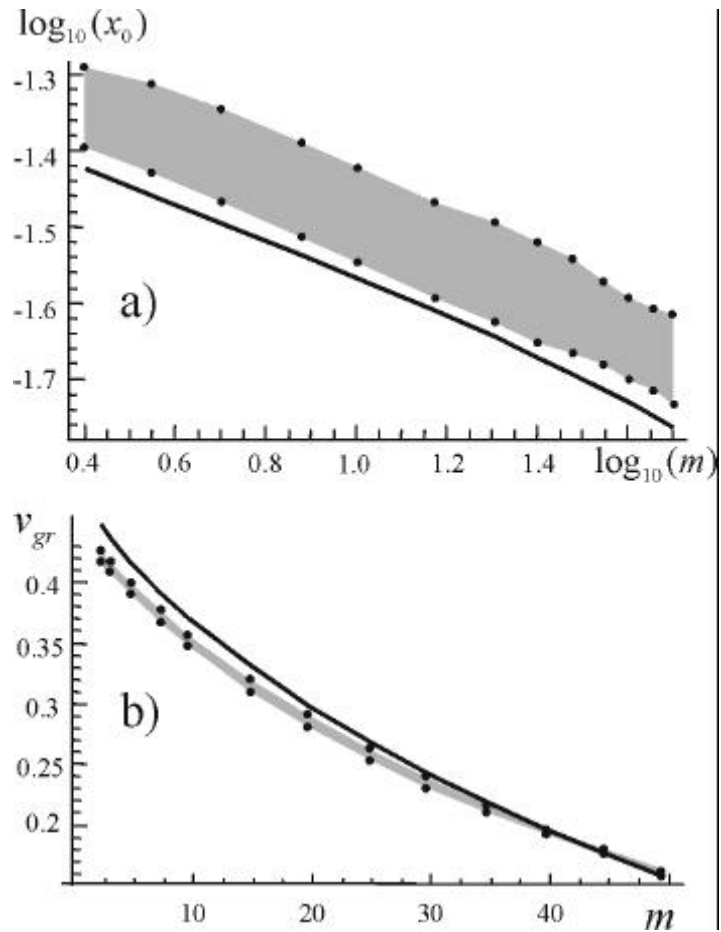

Fig. 3. The width (a) and the group velocity (b) of the mid-band DSS, as calculated from numerical integration of (1) (the solid circles correspond to marginal values at the boundaries of stability region), and as evaluated by linear theory (2)-(4). The parameters for numerical integration - as in Fig.1.

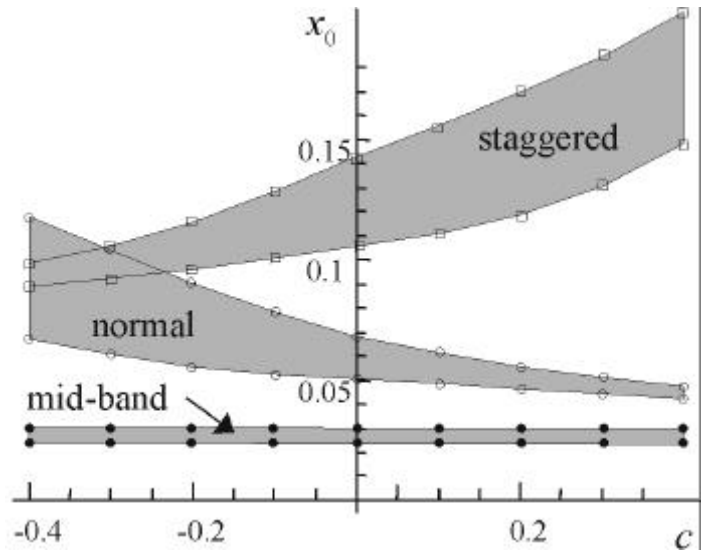

Fig.4. The width of the normal, staggered, and band-middle DSS, depending on focusing/defocusing parameter $c$, as calculated from numerical integration of (5) The solid circles, open circles, and squares correspond to the marginal values at the boundaries of stability region for mid-band, normal and staggered DSSs respectively. The parameters - as in Fig.1.